\documentclass{elsart}%
\usepackage{amssymb}
\usepackage{amsmath}%
\usepackage{amsfonts}%
\usepackage{graphicx}

\journal{Physics Letters A}

\begin{document}
\begin{frontmatter}
\title{Entanglement Enhanced Multiplayer Quantum Games \thanksref{publish}}
\thanks[publish]{{\em Physics Letters A} {\bf 302}, 229-233 (2002).}
\author[lqcc,mphy,bel]{Jiangfeng Du},
\ead{djf@ustc.edu.cn}
\author[mphy]{Hui Li},
\author[mphy]{Xiaodong Xu},
\author[mphy]{Xianyi Zhou}, and
\author[mphy]{Rongdian Han}
\address[lqcc]{Laboratory of Quantum Communication and Quantum Computation, University of
Science and Technology of China, Hefei, 230026, P.R.China.}
\address[mphy]{Department of Modern Physics, University of Science and Technology of China,
Hefei, 230027, P.R.China.}
\address[bel]{Service de Physique Th\'{e}orique CP225, Universit\'{e} Libre de Bruxelles,
1050 Brussels, Belgium.}
\begin{abstract}
We investigate the 3-player quantum Prisoner's Dilemma with a
certain strategic space, a particular Nash equilibrium that can remove the
original dilemma is found. Based on this equilibrium, we show that the game is enhanced
by the entanglement of its initial state.
\end{abstract}
\end{frontmatter}

\section*{Introduction}

Game theory is a very useful and important branch of mathematics due to its
broad applications in economics, social science and biology\cite{1}. Recently,
physicists specializing in quantum information theory are very interested in
the investigation of extending classical games into the quantum domain\cite{2}%
. Current researches have led new sights into the nature of
information\cite{3,3-1,3-2}, opening a new range of potential
applications\cite{6,6-1,7,8,9,10,11,12,13,13-1}. Quantum games are now found
helpful in developing new quantum algorithms\cite{6,6-1}, and quantum
strategies are found helpful in solving problems in classical
games\cite{7,8,9}. Investigations on evolutionary procedure of quantum games
are also presented\cite{10}. Although quantum games are mostly explored
theoretically, the quantum Prisoners' Dilemma has just been successfully
realized experimentally\cite{11}.

All the above works focus on 2-player quantum games (within a 2-qubit system).
However, S. C. Benjamin and P. M. Hayden recently presented the first study of
quantum games with more than two players\cite{13,13-1}. They showed that such
games can exhibit certain forms of pure quantum equilibrium that have no
analog in classical games, or even in 2-player quantum games. They also
proposed a physical model which is suitable for multiplayer quantum game, and
based their work on maximally entangled states. In this paper, we quantize the
3-player version of the Prisoners' Dilemma\cite{15}, which is a famous
multiplayer game in classical game theory, basing on the physical model
proposed in Ref. \cite{13,13-1}. We find a Nash equilibrium that can remove
the dilemma in the classical game when the game's state is maximally
entangled. This particular Nash equilibrium remains to be a Nash equilibrium
even for the non-maximally entangled cases. Furthermore, the payoffs at this
equilibrium increase monotonously when the amount of the game's entanglement
increases. Thus the quantum 3-player Prisoners' Dilemma is enhanced by the
entanglement of its initial state.

\section{3-Player Prisoner's Dilemma}

The Prisoner's Dilemma is a famous, none zero-sum game which illustrates a
conflict between individual and group rationality. The dilemma in this game
was first proposed by Merrill Flood and Melvin Dresher in 1950. Later on,
Albert Tucker made ``the Prisoner's Dilemma''\ as the title of this game who
wants the game to be more popular. After it was published, the dilemma
attracted widespread attention in a variety of disciplines. By far, 2-player
Prisoner's Dilemma has drawn interests of physicists specializing in quantum
information theory. The investigation of the 2-player quantum Prisoners'
Dilemma is presented\cite{2}. In this paper, we generalize this game to
involve 3 players, and investigate the quantization in a 3-qubit system.%
\begin{figure}
[ptb]
\begin{center}
\includegraphics[
height=1.1087in,
width=3.2482in
]%
{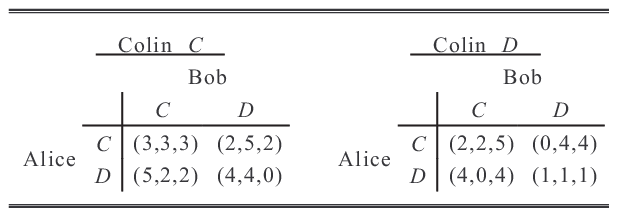}%
\caption{The payoff matrix of the 3-player Prisoners' Dilemma. The first entry
in the parenthesis denotes the payoff of Alice, the second number denotes the
payoff of Bob, and the third number denotes the payoff of Colin --- see in
Ref\cite{12}.}%
\label{Fig1}%
\end{center}
\end{figure}

The scenario of the 3-player Prisoner's Dilemma is as the following
hypothetical situation: three players --- Alice, Bob and Colin were arrested
under the suspicion of robbing the bank. They were placed in isolated cells
without a chance to communicate with each other. However, the police does not
have sufficient proof to have them convicted. Since each of them cared much
more about their personal freedom than about the welfare of their accomplice,
a clever policeman makes the following offer to the three players: Each of
them may choose to confess or remain silent. For convenience, we denote
choosing to confess by strategy $D$ (defect), and choosing to remain silent by
strategy $C$ (cooperate). If they all choose $D$ (defect), each of them will
get payoff $1$; if the players all resort to strategy $C$ (cooperate), each of
them will get payoff $3$; if one of the players choose $D$ but the other two
choose $C$, $5$ is the payoff for the former and $2$ for the latter two; if
one of the players choose $C$ while the other two adopt $D$, $0$ is payoff for
the former and $4$ for the latter two. This situation and its different
outcomes can be summarized by Fig. \ref{Fig1}. We can see that strategy $D$ is
the dominant strategy for each player, \textit{i.e.} whatever other players
do, each is better off defect than cooperate. In terms of game theory,
$(D,D,D)$ is the unique Nash equilibrium of the game, with payoff $(1,1,1)$.
Unilateral deviation from this equilibrium will decrease individual payoff.
Since each of them is a completely rational player, the game will definitely
end in the situation that all players choose strategy $D$. However, we can see
that $(C,C,C)$ can yield payoff $(3,3,3)$, much higher than $(1,1,1)$ in the
situation of $(D,D,D)$. In game theory, $(C,C,C)$ is a \textit{Pareto
Optimal}, which is regarded as the most efficient strategic profile of a game.
Unfortunately, rational reasoning forces each player to defect, and therefore
the game will end in $(D,D,D)$ rather than $(C,C,C)$. This is an instance of
that optimizing the outcome for a subsystem will in general not optimizing the
outcome for the system as a whole, and is exactly what called a dilemma in
this game.

\section{Physical Model for the Quantum Game}

\begin{figure}
[tb]
\begin{center}
\includegraphics[
height=0.9677in,
width=2.0159in
]%
{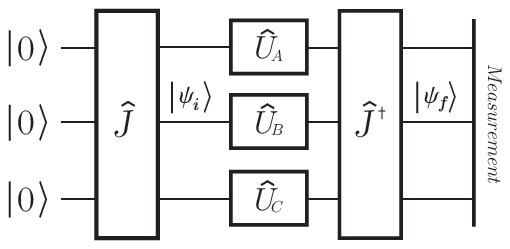}%
\caption{The physics model of a 3-player quantum game.}%
\label{Fig2}%
\end{center}
\end{figure}

In this paper, we make use of the physical model proposed in Ref.
\cite{13,13-1} --- see in Fig. \ref{Fig2}. In the board of\- this model, each
player is sent a 2-state quantum system (a qubit) and they can locally
manipulate their individual qubit. The possible outcomes of the classical
strategies \textquotedblleft Cooperate\textquotedblright\ and
\textit{\textquotedblleft}Defect\textquotedblright\ are assigned to two bases
$\left\vert 0\right\rangle $ and $\left\vert 1\right\rangle $ in the Hilbert
space of a single qubit. The state of the game is then described by a state in
the tensor product space of the three qubits, spanned by the bases $\left\vert
\sigma\right\rangle \left\vert \sigma^{^{\prime}}\right\rangle \left\vert
\sigma^{^{\prime\prime}}\right\rangle =\left\vert \sigma\sigma^{^{\prime}%
}\sigma^{^{\prime\prime}}\right\rangle $ ($\sigma,\sigma^{^{\prime}}%
,\sigma^{^{\prime\prime}}\in\left\{  0,1\right\}  $), where the first, second
and third entries refer to Alice's, Bob's and Colin's qubits respectively. The
initial state of the game is denoted by $\left\vert \psi_{i}\right\rangle
=\hat{J}\left\vert 000\right\rangle $, where $\hat{J}=\exp\left\{
i\frac{\gamma}{2}\hat{\sigma}_{x}\otimes\hat{\sigma}_{x}\otimes\hat{\sigma
}_{x}\right\}  $ $\left(  0\leqslant\gamma\leqslant\pi/2\right)  $ is the
entangling gate of the game and is known to all of the players. In fact,
$\gamma$ could be considered as a measure for the game's
entanglement.\ Strategic move of Alice (Bob or Colin) is denoted by unitary
operator $\hat{U}_{A}$ ($\hat{U}_{B}$ or $\hat{U}_{C}$), which are chosen from
a certain strategic space $S$. Since the strategic moves of different players
are independent, one player's operators only operate on his/her individual
qubit. Therefore the strategic space $S$ should be some subset of the group of
unitary $2\times2$ matrices. We can set the strategic space to be the
following 2-parameter subset of $SU(2)$ (for the choose of different strategic
spaces, see Ref. \cite{18,19,20}),%
\begin{equation}
\hat{U}\left(  \theta,\varphi\right)  =\left(
\begin{array}
[c]{cc}%
\cos\theta/2 & e^{i\varphi}\sin\theta/2\\
-e^{-i\varphi}\sin\theta/2 & \cos\theta/2
\end{array}
\right) \label{eq 1}%
\end{equation}
with $0\leqslant\theta\leqslant\pi$ and $0\leqslant\varphi\leqslant\pi/2$. To
be specific, $\hat{U}\left(  0,0\right)  $ is the identity operator $\hat{I}$
which corresponds to \textquotedblleft Cooperate\textquotedblright, and
$\hat{U}\left(  \pi,\pi/2\right)  =i\hat{\sigma}_{x}$, which is equivalent to
the bit-flipping operator, corresponds to \textquotedblleft
Defect\textquotedblright. It is easy to check that $\hat{J}$ commutes with any
operator consisting of $i\hat{\sigma}_{x}$ and $\hat{I}$ acting on different
qubits, hence the classical Prisoners' Dilemma could be faithfully recovered
by the quantum game.%

Having executed their strategic moves, the players forward their qubits to
gate $\hat{J}^{+}$, after which the game's final state prior to the
measurement can be written as%
\begin{align}
\left\vert \psi_{f}\right\rangle  & =\left\vert \psi_{f}\left(  \hat{U}%
_{A},\hat{U}_{B},\hat{U}_{C}\right)  \right\rangle \nonumber\\
& =\hat{J}^{+}\left(  \hat{U}_{A}\otimes\hat{U}_{B}\otimes\hat{U}_{C}\right)
\hat{J}\left\vert 000\right\rangle .\label{eq 2}%
\end{align}
The succeeding measurement yields a particular result with a certain
probability. Therefore the payoff for Alice should be the \textit{expected}
payoff%
\begin{align}
\$_{A}  & =5P_{DCC}+4\left(  P_{DDC}+P_{DCD}\right)  +3P_{CCC}\nonumber\\
& +2\left(  P_{CCD}+P_{CDC}\right)  +1P_{DDD}+0P_{CDD}\label{eq 3}%
\end{align}
where $P_{\sigma\sigma^{^{\prime}}\sigma^{^{\prime\prime}}}=\left\vert
\left\langle \sigma\sigma^{^{\prime}}\sigma^{^{\prime\prime}}\right\vert
\left.  \psi_{f}\right\rangle \right\vert ^{2}$ is the probability that
$\left\vert \psi_{f}\right\rangle $ collapses into $\left\vert \sigma
\sigma^{^{\prime}}\sigma^{^{\prime\prime}}\right\rangle $. Payoff functions of
Bob and Colin can be obtained from similar analyzing.

\section{Nash Equilibrium In The Quantum Game}

We now turn our attention to the Nash equilibrium of the 3-player quantum
Prisoner's Dilemma.

If the measure of the game's entanglement is $\gamma=0$, the game is
separable, \textit{i.e.} at each instance the state of the game is separable.
We find that all the eight strategic profiles consisting of $\hat{U}\left(
\pi,\pi/2\right)  =i\hat{\sigma}_{x}$ and $\hat{U}\left(  \pi,0\right)
=i\hat{\sigma}_{y}$ are Nash equilibria. However this situation of multiple
equilibria is trivial. For any profile of Nash equilibrium of the separable
game, because $i\hat{\sigma}_{x}\left\vert 0\right\rangle =i\left\vert
1\right\rangle $ and $i\hat{\sigma}_{y}\left\vert 0\right\rangle =-\left\vert
1\right\rangle $, the final state $\left\vert \psi_{f}\right\rangle =-\left(
-i\right)  ^{n}\left\vert 1\right\rangle \left\vert 1\right\rangle \left\vert
1\right\rangle $, where $n$ denotes the number of players who adopts $\hat
{U}\left(  \pi,\pi/2\right)  =i\hat{\sigma}_{x}$. According to the payoff
functions, each player receives payoff $1$. Hence $i\hat{\sigma}_{x}$ and
$i\hat{\sigma}_{y}$ have the same effect to the payoffs. Therefore all these
Nash Equilibria are equivalent to the classical strategic profile $(D,D,D)$.
Indeed, any quantum strategy $\hat{U}\left(  \theta,\varphi\right)  $ is
equivalent to the classical mixed strategy \textquotedblleft$C$ with
probability $\cos^{2}\theta/2$ and $D$ with probability $\sin^{2}\theta
/2$\textquotedblright, because $\hat{U}\left(  \theta,\varphi\right)
\left\vert 0\right\rangle =\cos\theta/2\left\vert 0\right\rangle
-e^{-i\varphi}\sin\theta/2\left\vert 1\right\rangle $ and hence the
measurement gives $\left\vert 0\right\rangle $ with probability $\cos
^{2}\theta/2$ and $\left\vert 1\right\rangle $ with probability $\sin
^{2}\theta/2$. Therefore we conclude that the separable quantum game does not
exceed the classical game.

Although the separable quantum game does not exhibit any quantum advantages,
the maximally entangled game does. It can be proved that the strategic profile
$i\hat{\sigma}_{x}\otimes i\hat{\sigma}_{x}\otimes i\hat{\sigma}_{x}$ is no
longer the Nash equilibrium. In fact,%
\begin{equation}
\$_{A}\left(  \hat{U}\left(  \theta,\varphi\right)  ,i\hat{\sigma}_{x}%
,i\hat{\sigma}_{x}\right)  =\left(  1+2\cos^{2}\varphi\right)  \sin^{2}%
\frac{\theta}{2}\leqslant3=\$_{A}\left(  i\hat{\sigma}_{y},i\hat{\sigma}%
_{x},i\hat{\sigma}_{x}\right)  \text{.}%
\end{equation}
Hence $i\hat{\sigma}_{x}$ is no longer the best response for one player when
the other two choose $i\hat{\sigma}_{x}$. Besides, we can also get that%
\begin{equation}
\$_{A}\left(  \hat{U}\left(  \theta,\varphi\right)  ,i\hat{\sigma}_{x}%
,i\hat{\sigma}_{y}\right)  =\frac{1}{2}\left[  7+3\cos\theta-2\sin^{2}%
\frac{\theta}{2}\cos2\varphi\right]  \leqslant5=\$_{A}\left(  \hat{I}%
,i\hat{\sigma}_{x},i\hat{\sigma}_{y}\right)  \text{.}%
\end{equation}
Therefore, considering the symmetry of the game, we obtain that among the
eight Nash equilibria in the separable game (consisting of $i\hat{\sigma}_{x}
$ and $i\hat{\sigma}_{y}$), any one containing $i\hat{\sigma}_{x}$ is not a
Nash equilibrium for the maximally entangled game.

However, the particular one $i\hat{\sigma}_{y}\otimes i\hat{\sigma}_{y}\otimes
i\hat{\sigma}_{y}$ remains to be a Nash equilibrium with
\begin{equation}
\$_{A}\left(  i\hat{\sigma}_{y},i\hat{\sigma}_{y},i\hat{\sigma}_{y}\right)
=\$_{B}\left(  i\hat{\sigma}_{y},i\hat{\sigma}_{y},i\hat{\sigma}_{y}\right)
=\$_{C}\left(  i\hat{\sigma}_{y},i\hat{\sigma}_{y},i\hat{\sigma}_{y}\right)
=3\text{.}\label{eq 7}%
\end{equation}
Indeed, for $\gamma=\pi/2$, with equation (\ref{eq 3}) and equation
(\ref{eq 7}), we have
\begin{equation}
\$_{A}\left(  \hat{U}\left(  \theta,\varphi\right)  ,i\hat{\sigma}_{y}%
,i\hat{\sigma}_{y}\right)  =\left(  1+2\cos^{2}\varphi\right)  \sin^{2}%
\frac{\theta}{2}\leqslant3=\$_{A}\left(  i\hat{\sigma}_{y},i\hat{\sigma}%
_{y},i\hat{\sigma}_{y}\right) \label{eq 8}%
\end{equation}
for all $\theta\in\left[  0,\pi\right]  $ and $\varphi\in\left[
0,\pi/2\right]  $. Analogously%

\begin{align}
\$_{B}\left(  i\hat{\sigma}_{y},\hat{U}\left(  \theta,\varphi\right)
,i\hat{\sigma}_{y}\right)   & \leqslant\$_{B}\left(  i\hat{\sigma}_{y}%
,i\hat{\sigma}_{y},i\hat{\sigma}_{y}\right) \nonumber\\
\$_{C}\left(  i\hat{\sigma}_{y},i\hat{\sigma}_{y},\hat{U}\left(
\theta,\varphi\right)  \right)   & \leqslant\$_{C}\left(  i\hat{\sigma}%
_{y},i\hat{\sigma}_{y},i\hat{\sigma}_{y}\right)
\end{align}
Hence, no player can improve his individual payoff by unilaterally deviating
from strategy $i\hat{\sigma}_{y}$, \textit{i.e.} $\left(  i\hat{\sigma}%
_{y},i\hat{\sigma}_{y},i\hat{\sigma}_{y}\right)  $ is a Nash equilibrium. It
is interesting to see that the payoffs for the players are $\$_{A}%
=\$_{B}=\$_{C}=3$, which is the best payoffs for them while remaining the
symmetry of the game. So the strategic profile $(i\hat{\sigma}_{y}%
,i\hat{\sigma}_{y},i\hat{\sigma}_{y})$ has the property of Pareto Optimal. By
allowing the players to adopt quantum strategies, the dilemma in the classical
game is completely removed when the game is maximally entangled.

\section{The Situation of Non-maximal Entanglement}%

\begin{figure}
[tb]
\begin{center}
\includegraphics[
height=1.7158in,
width=2.4137in
]%
{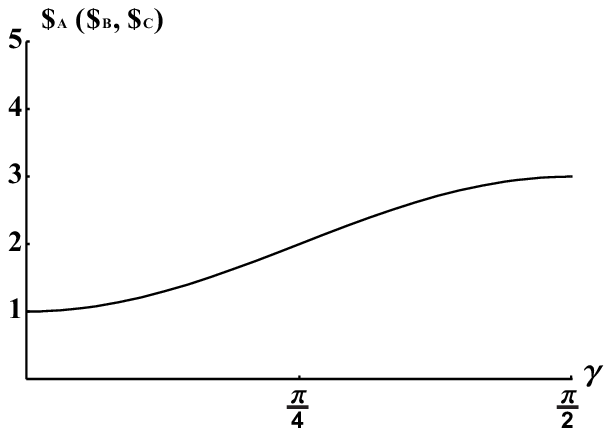}%
\caption{The payoff plot as a function of $\gamma$ when all the players resort
to Nash Equilibrium, $i\sigma_{y}\otimes i\sigma_{y}\otimes i\sigma_{y}$. From
this figure we can see that the payoffs of the players are the same
monotonuous increasing function of $\gamma$.}%
\label{Fig3}%
\end{center}
\end{figure}

It is well known that entanglement plays an important role in quantum
information processing and is viewed as the essential resource for
transmitting quantum information. In the preceding section, we have
investigated the maximally entangled game. In that case, we find a particular
Nash equilibrium $\left(  i\hat{\sigma}_{y},i\hat{\sigma}_{y},i\hat{\sigma
}_{y}\right)  $, which has the property of Pareto optimal. Because of the key
role of entanglement in quantum information, it will be interesting to
investigate how the properties of the quantum game relate to its entanglement.

The novel feature of $i\hat{\sigma}_{y}\otimes i\hat{\sigma}_{y}\otimes
i\hat{\sigma}_{y}$ is that this strategic profile remains to be a Nash
equilibrium for any $\gamma\in\left[  0,\pi/2\right]  $ . The proof runs as
follows. Assume Bob and Colin adopt $i\hat{\sigma}_{y}$ as their strategies,
the payoff function of Alice with respect to her strategy $\hat{U}\left(
\theta,\varphi\right)  $ is
\begin{align}
\$_{A}\left(  \hat{U}\left(  \theta,\varphi\right)  ,i\hat{\sigma}_{y}%
,i\hat{\sigma}_{y}\right)   & =\left(  1+2\cos^{2}\varphi\sin^{2}%
\gamma\right)  \sin^{2}\frac{\theta}{2}\nonumber\\
& \leqslant1+2\sin^{2}\gamma=\$_{A}\left(  i\hat{\sigma}_{y},i\hat{\sigma}%
_{y},i\hat{\sigma}_{y}\right)  \text{.}%
\end{align}
Hence, $i\hat{\sigma}_{y}$ is her best reply provided that her opponents all
choose $i\hat{\sigma}_{y}$. Since the game is symmetric, the same holds for
Bob and Colin. Therefore, no matter what the amount of the game's entanglement
is, $i\hat{\sigma}_{y}\otimes i\hat{\sigma}_{y}\otimes i\hat{\sigma}_{y}$ is
always a Nash equilibrium for the game. It can be seen that the payoff of the
players is a monotonously increasing function with respect to amount of the
entanglement,%
\begin{equation}
\$_{A}\left(  i\hat{\sigma}_{y},i\hat{\sigma}_{y},i\hat{\sigma}_{y}\right)
=\$_{B}\left(  i\hat{\sigma}_{y},i\hat{\sigma}_{y},i\hat{\sigma}_{y}\right)
=\$_{C}\left(  i\hat{\sigma}_{y},i\hat{\sigma}_{y},i\hat{\sigma}_{y}\right)
=1+2\sin^{2}\gamma\label{eq 6}%
\end{equation}

Fig. \ref{Fig3} illustrates how the payoffs depend on the amount of
entanglement when the players all resort to Nash equilibrium. From this
figure, we can see that entanglement dominates and enhances the property of
the game: the payoffs of the players are the same monotonously increasing
function of the amount of the game's entanglement. The strategic profile
$i\hat{\sigma}_{y}\otimes i\hat{\sigma}_{y}\otimes i\hat{\sigma}_{y}$ is
always a Nash equilibrium of the game no matter what\ the entanglement is. The
dilemma could be completely removed when the measure of game's entanglement
$\gamma$ increases to its maximum $\pi/2$.

\section{Conclusion}

As multipartite physical systems tend to be complex, multiplayer quantum games
may be more complicated and interesting than 2-player games\cite{12}. In this
paper, we present a symmetric quantum Nash equilibrium, $i\hat{\sigma}%
_{y}\otimes i\hat{\sigma}_{y}\otimes i\hat{\sigma}_{y}$, for the quantum
3-player Prisoners' Dilemma with certain strategic space. The novel feature of
this equilibrium is that it is the only surviving Nash equilibrium among those
for the separable game when the game is maximally entangled, and is the Pareto
optimal at the same time. The dilemma in the classical game could be
completely removed if all the players resort to this equilibrium. What is more
interesting is that it remains to be a Nash equilibrium whatever the amount of
entanglement is, and the payoffs for the players increase monotonously as the
amount of entanglement increases. It seems that in a multiplayer quantum game,
not only quantum strategies have superior performance over its classical
counterpart, but also entanglement can enhance the property of the game.
However we should point out that, in this paper we do not present other Nash
equilibria (if exist) for the quantum 3-player Prisoners' Dilemma, besides
$i\hat{\sigma}_{y}\otimes i\hat{\sigma}_{y}\otimes i\hat{\sigma}_{y}$. Finding
all the Nash equilibrium is an interesting but a little complicated task, and
deserves further investigation in future works. Another thing we should point
out is that our work is based on a restricted strategic space. As stated in
Ref. \cite{7,18}, the most general strategic space of the players could be all
of the trace-preserving, completely-positive maps. With this strategic space,
it has been demonstrated that if the amount of entanglement reaches it
maximum, there would be no pure strategic Nash Equilibrium in the 2-player
Prisoner's Dilemma\cite{18}. While maximally entangled multiplayer quantum
games can have certain forms of pure quantum equilibrium that have no analog
in classical games, or even in 2-player quantum games\cite{13,13-1}.
Especially when the entanglement varies, the game may have multiple or even
asymmetric Nash equilibria, and possibly other fascinating properties similar
to that in Ref \cite{8}. What singularity do multiplayer quantum games have
remain open in this paper and is interesting for further investigation. We
hope these works could also contribute to better understanding of multipartite
quantum information processing.

This project was supported by the National Nature Science Foundation of China
(Grants. No. 10075041 and No. 10075044) and Chinese Academy of Science.

\end{document}